\documentclass[
]{ceurart}

\usepackage{listings}
\begin{document}

\copyrightyear{2025}
\copyrightclause{Copyright for this paper by its authors.
  Use permitted under Creative Commons License Attribution 4.0
  International (CC BY 4.0).}

\conference{RobustIR@SIGIR25: The First Workshop on Robust Information Retrieval, July 17, 2025, Padua, Italy}

\title{A Research Vision for Web Search on Emerging Topics}

\author[1]{Alisa Rieger}[%
orcid=0000-0002-2274-1606,
email=alisa.rieger@gesis.org,
]
\cormark[1]
\address[1]{GESIS - Leibniz Institute for the Social Sciences, Germany} 

\author[1,2]{Stefan Dietze}[%
orcid=0009-0001-4364-9243,
email=stefan.dietze@gesis.org,
]
\address[2]{Heinrich Heine University Düsseldorf, Germany} 

\author[1]{Ran Yu}[%
orcid=0000-0002-1619-3164,
email=ran.yu@gesis.org,
]

\cortext[1]{Corresponding author.}

\begin{abstract}
  We regularly encounter information on novel, emerging topics for which the body of knowledge is still evolving, which can be linked, for instance, to current events. A primary way to learn more about such topics is through web search. However, information on emerging topics is sparse and evolves dynamically as knowledge grows, making it uncertain and variable in quality and trustworthiness and prone to deliberate or accidental manipulation, misinformation, and bias.
In this paper, we outline a research vision towards search systems and interfaces that support effective knowledge acquisition, awareness of the dynamic nature of topics, and responsible opinion formation among people searching the web for information on emerging topics.
To realize this vision, we propose three overarching research questions, aimed at understanding the status quo, determining requirements of systems aligned with our vision, and building these systems. 
For each of the three questions, we highlight relevant literature, including pointers on how they could be addressed. Lastly, we discuss the challenges that will potentially arise in pursuing the proposed vision.
\end{abstract}


\maketitle

\section{Introduction}
In recent years, we have witnessed an unprecedented transformation of our information ecosystem---arguably its most radical change in human history. The urgent need to understand the impact of this transformation on individuals and societies has been voiced by many~\cite{taylor_opinion_2018, trattner_responsible_2022, mitra_search_2025, shah_situating_2022, lorenz-spreen_systematic_2023}. 
Within the current information ecosystem, web search is one of the primary gateways to addressing a wide range of information needs, including searching for information on emerging topics. 
\textit{Emerging topics} are issues that are increasingly gaining attention in public discourse or within select communities, typically in response to recent or ongoing events or changes. They may emerge suddenly or evolve more gradually.
Because the body of knowledge around these topics is still developing, information is often scarce and changes rapidly, and different sources can provide divergent interpretations and viewpoints.
Thus, search engines are faced with two key challenges: (1) relevant content evolves quickly and tends to be diverse and inconsistent in quality, and (2) large-scale behavioral signals and structural data which search algorithms rely on, such as clickthrough rates and link graphs, are not yet available.
Today, LLM-based features, such as generated summaries, are increasingly integrated into dominant search engines, and many users turn to applications like \textit{ChatGPT} for search-related tasks. However, LLMs rely on large pretraining datasets that often lack timely content on recent developments, making them less effective at providing up-to-date information on emerging topics.
Hence, in the context of emerging topics, both information systems and users are susceptible to manipulation, misinformation, and viewpoint biases~\cite{rieger_responsible_2024, golebiewski_data_2019}.

\begin{figure*}
    \centering
\includegraphics[width=0.495\linewidth]{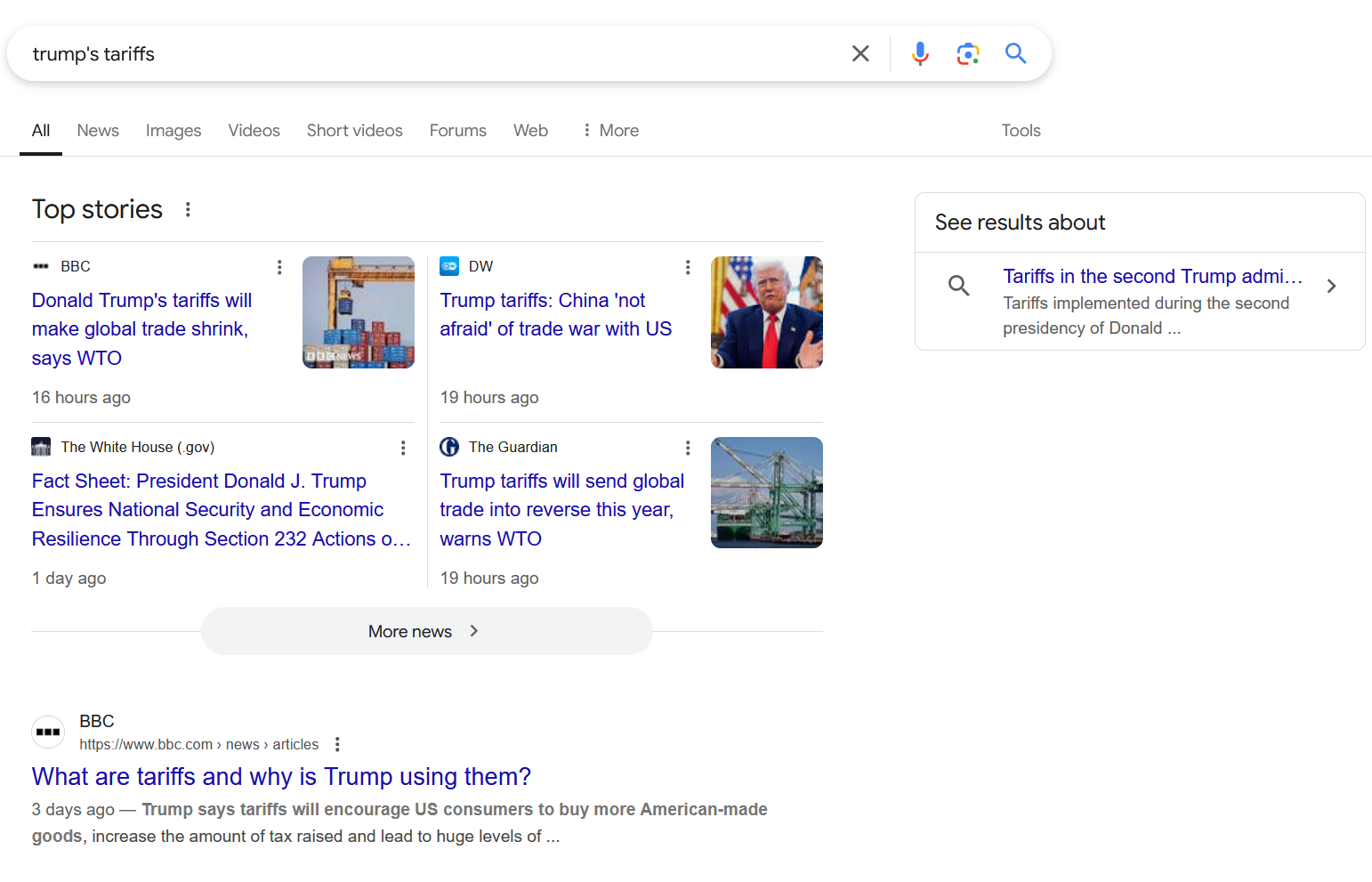}
\includegraphics[width=0.495\linewidth]{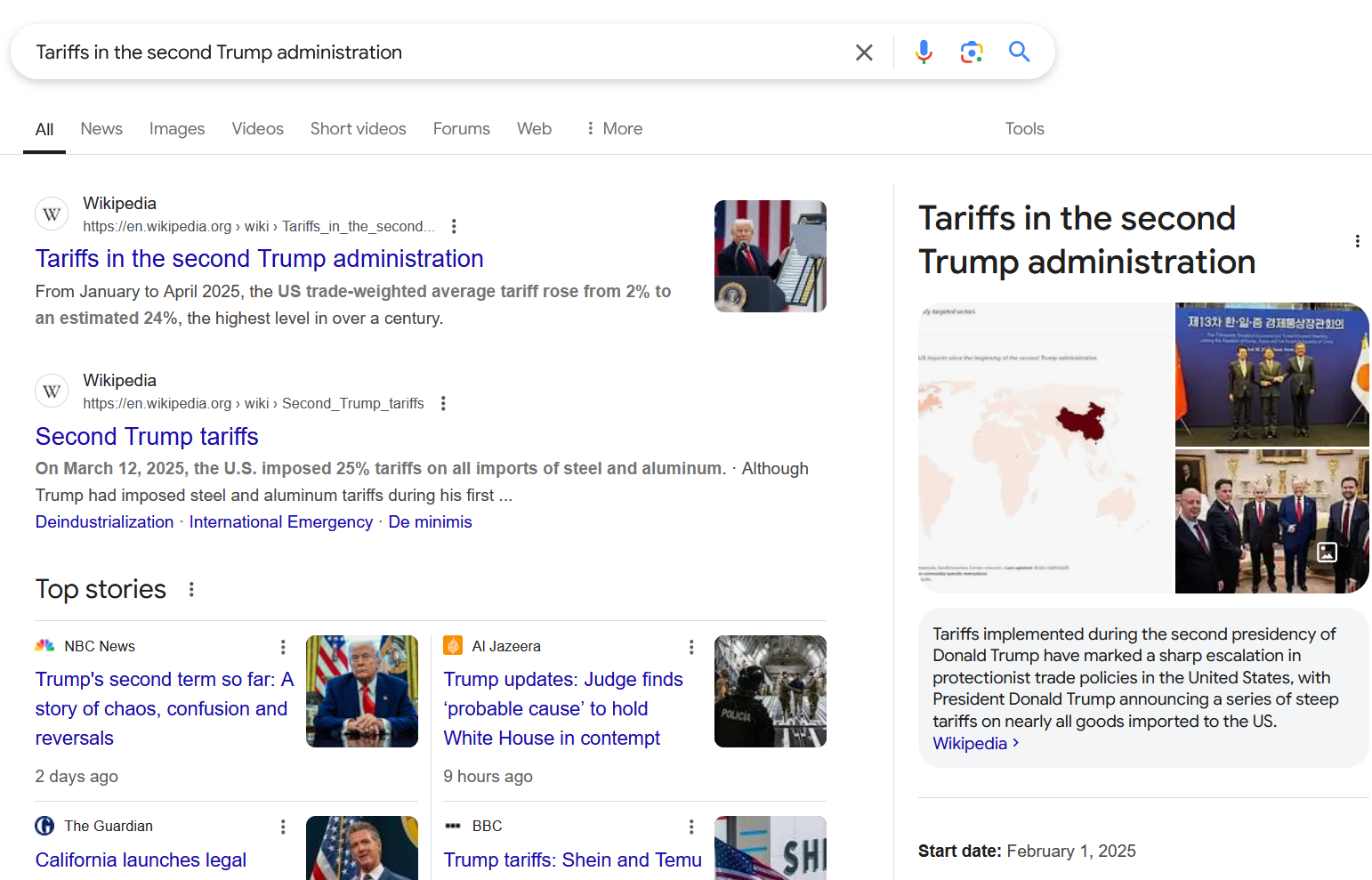}
    \caption{Search Page Example from Google after issuing the query \textit{trump's tariffs} (left) and after clicking \textit{see results about} feature (right), retrieved on 17 April 2025 }
    \label{fig:trumptrariffs}
\end{figure*}

To illustrate, let us consider a popular emerging topic at the time of writing this paper: \textit{Tariffs in the second Trump administration}. Engaging with information on this topic could have far reaching consequences for individuals and society, impacting not only someone's knowledge and opinions, but also their personal investment decisions, and on a broader scale, international relations. 
When looking at the search results for a query on this topic (see Figure~\ref{fig:trumptrariffs}), we can see that the results feature news articles from reputable outlets and offer an option to refine the search. Once refined, the top results prioritize Wikipedia entries accompanied by a snippet of the article and followed by related news coverage. 
These interface features seem carefully designed and might, indeed, support diligent information interaction behavior. 
Yet, emerging topics can vary widely in various regards, such as their levels of popularity and controversy, their complexity, or the scope of consequences they can carry for people and society.
We lack a structured understanding of how different search engines handle diverse emerging topics and how that influences human information behavior and opinion formation.

With this paper, we propose a research vision aimed at advancing the understanding of web search on emerging topics by analyzing how current search engines handle such topics, as well as determining how to design search environments that support users in engaging diligently and effectively with information on emerging topics. To this end, we highlight key literature and methodologies to provide a foundation for research focused on both users and search systems.

\section{Sociotechnical Imaginary and Research Questions} 
\label{sec:sociotechnical_imaginary}

\begin{figure}
    \includegraphics[width=1\linewidth]{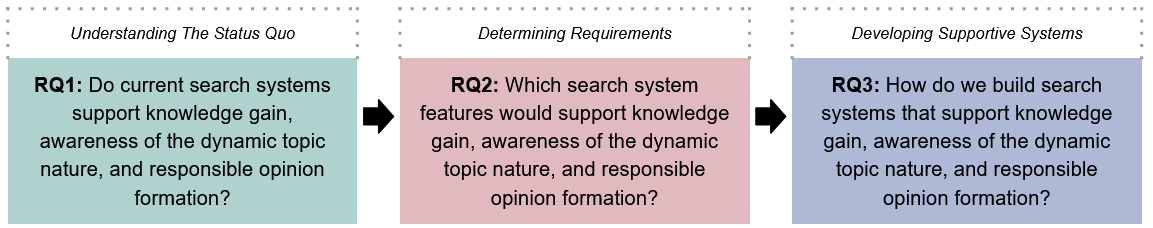}
    \caption{Overview of Main Research Questions}
    \label{fig:researchquestions}
\end{figure}

The research vision that we suggest with this paper is anchored in the  sociotechnical imaginary~\cite{mitra_search_2025} of a \textit{healthy and emancipatory information ecosystem} that prioritizes societal needs, such as an informed citizenry and the sustainable use of resources, and user needs, such as agency, transparency, and effective knowledge gain~\cite{shah_envisioning_2024}.
From this overarching imaginary, we derive the research objectives of designing search systems that support effective knowledge gain, awareness of the dynamic topic nature, and responsible opinion formation in people who search for information on emerging topics. 
While these objectives can provide a starting point, they should be refined iteratively, based on insights gained through system- and user-centered quantitative and qualitative research, theoretical analysis, and participatory design involving various stakeholders and experts.
To pursue this research vision, we suggest three main \textbf{research questions}, outlined in Figure~\ref{fig:researchquestions}. 

In addition to trying to identify approaches to support users and reduce potential harms within the search systems that currently exist, efforts should also be spent considering solutions beyond the systems we know and rely on. 
Alternative information environments with different characteristics might, in fact, be better suited to support information behavior that is beneficial for the individual and society in the context of emerging topics. 

\section{Positioning Within the Research Landscape}
In the following, we propose more detailed questions and highlight related work that may inform and support research efforts on each of the three main research questions that we outlined in Figure~\ref{fig:researchquestions}.

\subsection{Understanding the Status Quo (RQ1)}
\label{sec:statusquo}

To gain insights into the status quo of web search on emerging topics, we can consider the following guiding questions, among others: 
\begin{itemize}
    \item What is a meaningful and operational definition of emerging topics in the context of information retrieval research?
    \item How do different search engines handle and present information on emerging topics?
    \item How does the way search engines present information currently affect search behavior and search outcomes?
    \item With what information needs about emerging topics do people turn to web search?
\end{itemize}

Although we have an intuitive understanding of emerging topics as novel issues that are increasingly gaining attention, we lack a concrete operational definition. Developing such a definition is a critical first step for all subsequent research. For that, useful reference points can be found in research investigating how online information and user engagement develop for different topics over time, for example, to understand the life cycle of news stories~\cite{castillo_characterizing_2014} and the dynamics of online attention~\cite{lorenz-spreen_accelerating_2019}, or to detect trending topics and events~\cite{cataldi2014personalized, zubiaga_realtime_2015, hu_event_2022}.

Algorithmic systems run by private companies play a vital role in our information behavior, yet growing concerns have been raised about whether they ultimately support or undermine an informed citizenry and healthy democratic societies~\cite{lorenz-spreen_systematic_2023,taylor_opinion_2018, rieger_striving_2024, hills_dark_2019, trattner_responsible_2022}.
The increasing lack of transparency over such systems has sparked a growing need for \textit{algorithmic auditing} studies to investigate the functionalities and impact of different algorithmic systems~\cite{ulloa_scaling_2024}.
Auditing studies have been conducted, for instance, to investigate personalization and partisan bias for political queries and in the context of  elections~\cite{urman_matter_2022,robertson2018auditing, metaxa_search_2019, hu_auditing_2019}, gender bias in image search~\cite{otterbacher_competent_2017, singh_female_2020}, or viewpoint bias for queries on debated topics~\cite{draws_viewpoint_2023}.

To understand how search engines handle emerging topics, an auditing study by~\citet{trielli_search_2019} on news-related search functions offers methodological inspiration. With their study, \citet{trielli_search_2019} tried to understand how search engines shape the availability and consumption of news stories and which editorial news values (e.g., recency, relevance, social impact) might underlie the algorithmic filtering processes. Over one month, they applied a method to determine the most relevant news stories every day, generate search queries for these stories, use these queries in automated Google searches, and finally scrape and analyze the retrieved results, focusing on the \textit{Top Stories} box. The researchers found a preference for selected news sources and identified recency as an underlying editorial value. 
Similar studies could provide a systematic understanding of the functions (e.g., ranking criteria) and interface features (e.g., summaries) of different search engines in the context of emerging topics. 

Search engine functions and interface features have changed considerably and become more diverse in recent years~\cite{oliveira_evolution_2023}. Today, most search engines apply featured snippets and LLM-generated conversational search elements such as summaries for some, but not all queries.
Such features were found to have a substantial impact on human information behavior, exposure diversity, and opinion formation~\cite{bink_investigating_2023, sharma_generative_2024}. 
The evolution of search engine result pages for queries on emerging topics over time (i.e., from the initial emergence of a topic to it becoming more established), as well as their impact on search interactions and outcomes should be carefully evaluated.

To study information behavior and its outcomes in algorithmic environments, recent research has drawn on both quantitative and qualitative approaches.
In quantitative user studies, a wide variety of metrics have been employed to assess relevant search interactions and outcomes. 
With behavioral and questionnaire-based methods researchers have measured, for instance, search diligence~\cite{rieger_potential_2024}, search interaction bias~\cite{robertson_users_2023, rieger_disentangling_2024}, knowledge gain~\cite{otto2021predicting,yu2018predicting,gadiraju2018analyzing,yu2021topic, rieger_disentangling_2024, peng2021eye}, or opinion change~\cite{white_belief_2014, draws_this_2021, rieger_nudges_2024}. 
In light of the multidimensionality of web search on emerging topics, diverse combinations of task and user dimensions that can shape search interactions and outcomes need to be considered. 
Information needs related to an emerging topic will impact the task type, complexity, and goal. Additionally, individuals bring varying levels of prior topic knowledge and domain knowledge, as well as different attitudes and user traits.
Prior research exploring the ways in which search interactions and outcomes are shaped by characteristics of the task~\cite{liu_deconstructing_2021, hienert_role_2018, obrien_role_2020} and the user~\cite{white_characterizing_2009, al-samarraie_impact_2017, rieger_disentangling_2024} can provide a valuable foundation for similar research efforts in the context of emerging topics.

To gain a deeper and more contextualized understanding of the role of web search in people's broader information behavior on emerging topics across different platforms (e.g., social media, conversational AI) requires qualitative, in addition to quantitative research.
For that, methodological inspiration can be drawn from recent research by~\citet{hassoun_practicing_2023} or \citet{molem_keepin_2024} who conducted interview and diary studies to gain insights into information behavior and opinion formation in online environments.

\subsection{Determining Requirements (RQ2)}
For determining the system and interface requirements to support web search on emerging topics, the following questions may provide some guidance:
\begin{itemize}
    \item What search interface features would support and motivate diligent information behavior and awareness of the dynamic nature of emerging topics?
    \item  How should information on emerging topics be presented to promote effective knowledge gain and appropriate reliance?
    \item How do human experts (e.g., Journalists) curate and present information on emerging topics?
\end{itemize}

 Research in the fields of HCI and UX often revolves around understanding how to design systems and interfaces that assist users in achieving specific objectives efficiently.
In recent years, research focusing on societal needs, in addition to individual user needs, has started to receive more attention~\cite{mitra_search_2025, lorenz-spreen_how_2020, werthner_perspectives_2022}. 
This has been driven by a growing awareness of the central role digital technologies play in human information behavior and the profound societal impact they have.
In addition to analyzing and critiquing information technology through the lens of individual and societal needs, researchers have also explored interventions to mitigate risks and ways of creating alternative information environments that may be better aligned with these needs. 
For instance, researchers have suggested different tools to support truth and autonomy in online information behavior~\cite{lorenz-spreen_how_2020, kozyreva_toolbox_2024}.
Methodological insights for evaluating the impact of such interventions in web search on emerging topics can be drawn from studies investigating how they affect users' reliance, search interactions, and outcomes when searching for information on debated topics~\cite{bink_balancing_2024, rieger_potential_2024, rieger_nudges_2024, bink_investigating_2023}.

Participatory design methods involving diverse stakeholder groups, as used for search interface design, for instance, by~\citet{paramita_improving_2024}, could offer valuable insights into what search engine functions or alternative information environments people want and need for exploring emerging topics.
To ensure that research builds on established knowledge and practices, it is important to consult domain experts when determining system and interface requirements for web search on emerging topics. 
Journalists, who navigate the challenges of presenting information on emerging topics on a daily basis, can likely provide valuable insights.

\subsection{Developing Supportive Systems (RQ3)}

Building on the first two research questions, the third focuses on addressing the potential disconnect between what existing systems offer and what users and society need.
Although its specifics depend on insights from the earlier questions, the following questions will almost certainly need to be addressed:

\begin{itemize}
    \item How can we detect emerging topics?
    \item How can we ensure relevance and quality of the retrieved information?
    \item What retrieval and ranking approaches are able to handle dynamically evolving and data-sparse information spaces?
    \item How can we create meaningful summaries that reflect the dynamic nature of information on emerging topics?
\end{itemize}

Ensuring that search engines can handle queries on emerging topics in a distinct manner requires methods to detect emerging topics early. 
For advancing the detection of emerging topics, researchers can draw on various approaches to detect novel topics and events in the context of web search, social media, news, multimedia and cross-platform data. 
In the context of search, prior work has investigated supervised classification, relying on sudden changes in query frequency~\cite{ren2013understanding}, leveraging contextual information from search logs, results, news, and blogs~\cite{zhang2018automatic,ren2013understanding,sun2011query}, incorporating user feedback \cite{diaz2009integration,diaz2009adaptation}, or clustering related queries \cite{louis2011use}. 
For novel topic and event detection in social media, researchers have explored classification with advanced NLP techniques \cite{deviatkin2018discovering}, (incremental) clustering \cite{li2017real,chen2013emerging}, topic models \cite{wang2015detecting}, methods combining term aging with social relationships \cite{cataldi2010emerging,cataldi2014personalized}, and entity extraction and topic relation analysis~\cite{cataldi2010emerging}.
For news, multimedia, and cross-platform data, prior research has investigated topic modeling and clustering~\cite{bao2015cross, radinsky2013mining}, keyword frequency analysis \cite{kasiviswanathan2011emerging}, and dictionary learning~\cite{kasiviswanathan2012online,kasiviswanathan2013novel}.

For supporting effective knowledge gain, it is essential that retrieved resources are both relevant and of high quality. Yet, for emerging topics, conventional quality signals such as user engagement metrics and historical link structures might not yet be available. This calls for ranking approaches that can assess content reliability and relevance even in the absence of rich behavioral or historical data.
To this end, methods for misinformation detection based on pattern-based approaches might be useful, such as studies that examined writing styles and found that news titles help distinguish fake from real content \cite{horne2017just}. In this line of work, both supervised \cite{shu2020hierarchical} and unsupervised \cite{ghosal2020resco} models, as well as deep learning methods \cite{guacho2018semi,przybyla2020capturing,ghanem2021fakeflow}, have been used. 
However, new methods for assessing content quality may be needed in the context of emerging topics for which information might be incorrect due to their evolving nature rather than deliberate manipulation, and in the era of LLMs capable of generating plausible incorrect information.

To support users with limited prior topic knowledge in gaining knowledge on emerging topics, we need methods to create high-quality summaries, despite the challenge of sparse and fast-changing information. 
Earlier work on extractive summarization, which selects text snippets directly from source documents \cite{sarker2016query,rudra2015extracting}, abstractive summarization, which generates summaries without reusing original sentences \cite{nallapati2016abstractive,yang2019mares}, and event summarization, which incorporates timelines \cite{kedzie2016real,zopf2016sequential} can serve as a helpful foundation for creating such summaries. 
Further, recent advances in LLMs have shown promising results in generative summarization in different domains~\cite{pu2023summarization, zhang2024systematic}.
We also need to ensure that the summaries do not elicit a false sense of certainty about information that is inherently dynamic and thus uncertain.
When evaluating summarization techniques, we thus propose going beyond common performance metrics, by conducting user studies that measure their effect on search behavior and search outcomes, such as knowledge gain (see Section~\ref{sec:statusquo}).

\section{Anticipated Research Challenges}
A major challenge of researching web search on emerging topics lies in the temporal dimension of emerging topics. Studying system and user behavior around real emerging topics requires rapid study deployment and adaptable plug-in research designs when a new topic starts to emerge. This makes it difficult to rely on pre-processed search results, e.g., with quality or viewpoint labels which are often essential for deeper analysis and controlled manipulation of the search results users get presented with in user studies. Further, when a topic is identified as emerging, it might already have reached some popularity and researchers might miss out on investigating system and user behavior in the early stages of an emerging topic.

Another central challenge will be designing studies that capture real user behavior without sacrificing feasibility or interpretability.
This task is made more difficult by the increasing opacity of online information platforms, as tighter API restrictions restrictions hinder access to large-scale, real-world data needed to understand algorithmic effects.
Attempts to work around these limitations by collecting data with controlled study designs with tailored search interfaces and artificial search tasks and scenarios might, however, undermine ecological validity and limit the relevance of findings to real-world information behavior and systems.
For instance, recent research by~\citet{hassoun_practicing_2023} has shown that information journeys are becoming less linear and the role of search engines in these journeys is becoming less central. 
To avoid conducting research based on outdated assumptions, we need to stay up-to-date with the role that web search plays in the broader information behavior on emerging topics, and with evolving search engine functions~\cite{oliveira_evolution_2023}.

Timely detection requires integrating diverse sources (e.g. social medias) beyond the search engine, and accessing relevant data at scale can be difficult. Optimizing search and ranking algorithms involves trade-offs between factors like quality, bias, timeliness, and relevance. Their interplay, along with system objectives, must be carefully reflected in the design. Finally, testing and deployment at scale are hindered by researchers’ limited access to real-world platforms, computational resources, interactions, and data logs.
 
Based on the socio-technical imaginary and research objectives that we have proposed (see Section~\ref{sec:sociotechnical_imaginary}), it is clear that we need to consider multiple relevant metrics that capture search behavior, search outcomes, and user reflection.
In addition, the broad and varied nature of emerging topics gives rise to a wide range of possible user, task, and topic dynamics. 
To design manageable studies, researchers must make choices that narrow the combinations of users, tasks, topics, and metrics, which may limit the generalizability of findings~\cite{liu_deconstructing_2021, liu_cranfieldinspired_2022}.

\section{Concluding Remarks}
In recent years, our information ecosystem has undergone a radical transformation, with the full impact on individuals and societies remaining largely unclear. Researchers exploring the impact often come to concerning conclusions.
With this paper, we aim to motivate research into the design of information systems that align with individual and societal needs, specifically focusing on web search in the context of emerging, often controversial topics.
To that end, we outline a research vision towards gaining a better understanding of the status quo of system and user behavior and developing systems that support effective
knowledge gain, awareness of the dynamic topic nature, and responsible opinion formation. Ultimately, the envisioned research aims to contribute to a healthy and emancipatory information ecosystem.

\begin{acknowledgments}
  This work was partially funded by the DFG, German Research Foundation (``EmergentIR'', 548295069).
\end{acknowledgments}

\section*{Declaration on Generative AI}
 During the preparation of this work, the authors used Grammarly and ChatGPT in order to: Grammar and spelling check, Paraphrase and reword. After using the tool, the authors reviewed and edited the content as needed and take full responsibility for the publication’s content. 

\bibliography{references}

\end{document}